\documentclass[12pt]{article}
\usepackage{hyperref,graphicx,floatflt}
\begin{document}
\begin{center}
\textbf{\Large To the 90-th anniversary of\\[0.3cm] Dmitry Vasilievich Volkov's birthday}\\[0.3cm]
{\large A.S.~Bakai, S.V.~Peletminskii, N.F.~Shul'ga, Yu.V.~Slyusarenko,\\[0.2cm] D.V.~Uvarov, A.A.~Zheltukhin}\\[0.2cm]
\textit{NSC Kharkov Institute of Physics and Technology,}\\ \textit{61108 Kharkov, Ukraine}\\[0.5cm]
\end{center}
\begin{abstract}
The essay is devoted to the personality of the prominent  theorist  D.V.~Volkov  and his pioneer works in quantum field theory  and elementary particle physics.
\end{abstract}

\begin{floatingfigure}[l]{5cm}
\includegraphics[scale=0.4]{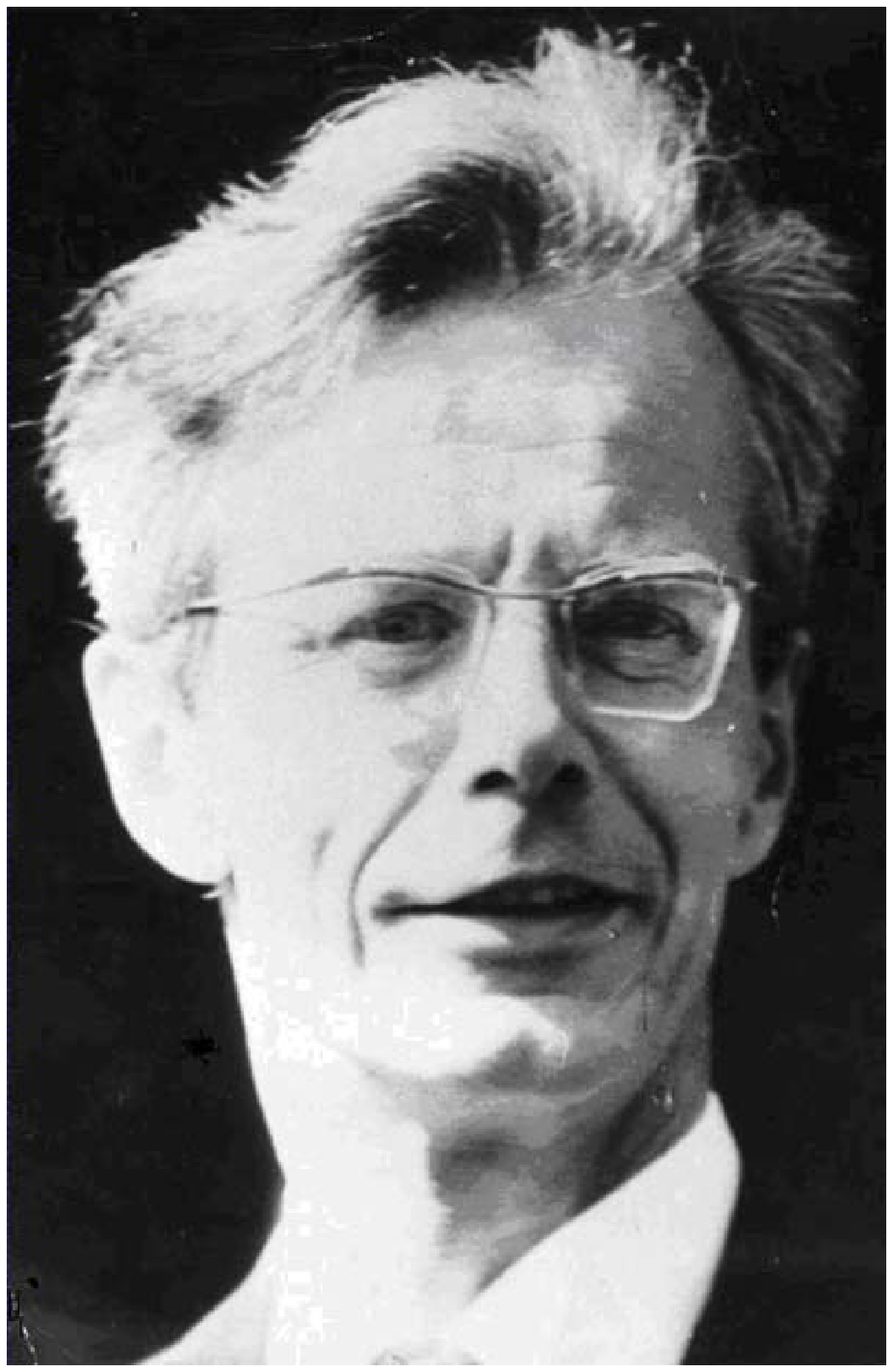}
\end{floatingfigure}
On July 3, 2015, ninety  years  have passed after the birthday of  Dmitry Vasilievich Volkov, an outstanding physicist,  Honoured Science Worker, Academician  of  National Academy of Sciences  of Ukraine, one of  pioneers of the theory of supersymmetry and supergravity, who made fundamental contribution to elementary particle physics  and quantum  field theory.  In his works, D.V.~Volkov laid the foundation of the new approach to the understanding of the space-time structure and the unification of quantum theory with general relativity.  After the discovery of the Higgs boson at CERN in 2013, the detection of new particles predicted by supersymmetry became one of the top priorities in the research program of the Large Hadron Collider.

D.V.~Volkov was born on July 3, 1925, in the city of Leningrad (now St. Petersburg).  His father Vasily Nikolaevich Volkov was a carpenter, his mother Olga Ivanovna Kazakova was a kindergarten teacher. In 1941, when the Great Patriotic War started, Vasily Nikolaevich joined the Leningrad volunteer corps and went missing in action. Dmitry's elder brother Lev, a military cadet, was fatally wounded in December of 1941 in Leningrad. During hard war years Dmitry worked at a collective farm, at a factory, and in 1943 as an eighteen-year-old youth was drafted into the Red Army and participated in the  battles at the Karelsky and the First Far East fronts. The war strengthened his character and formed his civic stand. After demobilization, he came back to Leningrad and passed examinations for 9-th and 10-th grades without attending lectures, and in 1947 entered the Physics Department of the Leningrad State University.  Dmitry studied during four years in Leningrad and, together with other excellent students, in 1951 he was transferred to the Kharkov State University, where at that time the Nuclear Physics Division was organized at the Physics and Mathematics Department.

After graduating from the university Dmitry entered into a doctoral program under the supervision of Prof. Aleksandr Ilich Akhiezer, who organized a group of talented postgraduate students for investigating the problems of quantum electrodynamics.  This group included V.F.~Aleksin, V.G.~Baryakhtar, P.I.~Fomin, S.V.~Peletminskii and D.V.~Volkov, who later became world-known scientists. The 1950's are known as a period of extensive development in the field of  nuclear and elementary particle physics and quantum electrodynamics. In 1956 D.V.~Volkov defended his Candidate of Science dissertation on scalar electrodynamics and was invited to work  at the Ukrainian Institute of Physics and Technology (now  National Science Centre Kharkov Institute of Physics and Technology), where he studied quantum field theory. A world recognition came to D.V.~Volkov in 1959 when he discovered the parastatistics (the Green-Volkov statistics), a new way of quantization of half-integer spin wave fields which generalized the  Bose-Einstein and the Fermi-Dirac statistics.

Shortly after, Dmitry Vasilievich began to develop an original approach to studying the Regge pole theory. In 1962, in cooperation with Vladimir Naumovich  Gribov,  D.V.~Volkov discovered the phenomenon of connection between Regge poles in the nucleon-nucleon and nucleon-antinucleon scattering amplitudes later  called "the Regge pole conspiracy". This "conspiracy" theorem gave rise to a great flood of theoretical and experimental investigations in high energy physics. At the same period D.V.~Volkov introduced an important notion of the collinear symmetry subgroups (simultaneously with G.~Lipkin and S.~Meshkov), that allowed to describe the scattering amplitudes  of particles and their classification on the basis of representations of the groups of higher symmetries.

In the middle of the 1960s Volkov concentrated his scientific activity on the development of the current algebra and spontaneously broken symmetries.  In 1968 he built a general approach to the description of the Nambu-Goldstone fields associated with an arbitrary spontaneously broken group of internal symmetry.  Simultaneously analogous results were obtained by J.~Wess, S.~Coleman, B.~Zumino and C.~Callan.  The Volkov's approach was based on  the deep insights into the group-theoretical methods of Elie Cartan and understanding of their universal role in the description of physical systems with degenerate vacua. Volkov's papers of this cycle gave a remarkable example of the unity of rigorous  mathematical methods and great physical intuition so typical for his creative approach.  In statistical physics Volkov applied his approach to the description of magnons as the Nambu-Goldstone particles, and  in 1970, together with his students,  Volkov built a general  phenomenological Lagrangian for spin waves in magnetically ordered and  disordered media. These papers became the basis for the novel approach that allowed to obtain new results in condensed matter theory.

At the same time Volkov started thinking over the problem of the existence of the Nambu-Goldstone fermions.  In 1971 at the International Seminar at the Lebedev Institute of Physics in Moscow he presented a new construction generalizing the concept of the spontaneously-broken internal symmetry groups to the case of groups of a new type including the Poincare group as a subgroup. This new symmetry, later called supersymmetry, allowed to overcome the well-known no-go theorem of Coleman and Mandula forbidding a non-trivial unification of internal and space-time symmetries.  The Volkov's construction proved the possibility of the existence of the Nambu-Goldstone fermions, and in 1972 Volkov, together with his student V.P.~Akulov built their phenomenological Lagrangian. A little bit earlier in 1971, supersymmetry was discovered by Yuri Golfand and Evgeny Likhtman and rediscovered in 1974 by J.~Wess and B.~Zumino.   The latter extended  the two-dimensional world-sheet graded Lie algebra discovered in 1971 in  dual models and string theory by  P.~Ramond,  and by J.~Schwarz, A.~Neveu,  and by D.~Gervais and B.~Sakita,  to four-dimensional  space-time.  Since then the ideas of supersymmetry and strings became the fundamental theoretical conceptions in the elementary particle and astroparticle physics.

 In 1973 Volkov advanced the idea of  the unification of supersymmetry with general relativity,  and  together with his student V.A.~Soroka constructed the first theory of supergravity based on the consideration of  the super-Poincare group as a gauge group.  This unification extended the Hilbert-Einstein theory, treated as a gauge theory of the boson field with spin 2 (the graviton), by the addition of the 3/2-spin Rarita-Schwinger fermionic gauge field (the gravitino) accompanied with the Nambu-Goldstone fermion.  They showed that spontaneous breakdown of the local supersymmetry in the presence of the Nambu-Goldstone fermion resulted in a supersymmetric extension of the Higgs effect to the gravitino.  This seminal paper stimulated the appearance of the papers made in 1976 by D.~Freedman, P.~van Nieuwenhuizen, S.~Ferrara,  and by S.~Deser and B.~Zumino,  where  pure supergravity including  only the graviton and  gravitino fields,  was described.

 During the 1970s Dmitry Vasilievich in collaboration with his students solved the complicated problem of spontaneous vacuum transitions in the dual resonance models of Veneziano and Neveu-Schwarz. It permitted to discover a hidden quark structure of the Regge trajectories accompanied with a new infinite-dimensional symmetry of the dual amplitudes.

Later, D.V.~Volkov proposed the mechanism of spontaneous compactificaton of redundant space dimensions in supersymmetric theories of gauge fields. In collaboration with his students Volkov built new models for interacting gauge and gravitational fields generalizing  the Kaluza-Klein model and invariant under the symmetries of the Standard Model.

In the late 1980s Volkov developed a new supertwistor approach in the supersymmetric theory of particles, strings and membranes,  and in collaboration with his students built new twistor-like models of these objects. Based on this, Volkov explained the mystery of the fermionic kappa-symmetry as   superdiffeomorphisms of world lines and world sheets of particles and strings. The progress achieved by using supertwistors stimulated an intense activity in many research groups in the world.

Volkov's works in the field of supersymmetry and the associated problems won a wide international recognition. They are cited as the basic ones for the present-day stage in the development of high-energy physics and field theory. In 1994 Volkov was invited as a guest of honor with the talk "Supergravity before 1976" at the international conference in Erice, Sicily, which was devoted to the history of original ideas and basic discoveries in particle physics of the twentieth century.

 In his last talk presented at the SUSY-95 conference in Paris, Volkov proposed a new generalized action principle for superstrings and supermembranes. Taking into account his great contribution to the development of elementary particle physics, the organizers devoted the Conference Proceedings to the memory of Dmitry Volkov.

The scientific activity of Dmitry Vasilievich was inseparably linked with Kharkov Institute of Physics and Technology where he worked for more than 40 years and created a scientific school known far outside Ukraine's borders. D.V.~Volkov made a lot of effort into scientific-organizational and administrative work. He was a member of the Scientific Committee on Nuclear Physics at the National Academy of Sciences of Ukraine, a member of the editorial board of the Soviet Journal of Nuclear Physics (Moscow), the journal "Problems of Nuclear Physics and Cosmic Rays"  published by V.N.~Karazin Kharkov National University.  For more than 30 years he was the head of the Library Council of KIPT, a member of many scientific boards, and the supervisor of the scientific seminar at the V.N.~Karazin KNU.

The achievements of D.V.~Volkov were acknowledged through orders and medals.  In 1997, posthumously, Dmitry Volkov was awarded the Walter Thirring Medal, and in 2009 the State Prize of Ukraine in Science and Technology (together with his students).

The colleagues and co-workers of Dmitry Vasilievich highly appreciated a great profoundity of his thinking, a subtle feeling of beauty in science, his ability to catch instantly the essence of the discussed problem and to react with original solutions.

Many people noted on his uncompromising attitude to any violation of scientific ethics, to careerism, bureaucracy in science and to injustice.  Dmitry Vasilievich was a modest person with good will and respectful attitude to people, readiness to help those in need.

Despite the heart disease which resulted from the hard war years, Volkov worked on the maximum of his physical forces, inspired by new ideas. Contacts and discussions with Dmitry Vasilievich on any problems of science and life gave enormous enjoyment, created optimism and belief in the triumph of wisdom and kindness.  The name of D.V.~Volkov, a knight of theoretical physics, forever remains in the history of science and in the memory of his colleagues and friends.

Additional information about life of D.V. Volkov and his outstanding contribution to science can be found in Refs.~\cite{DVbook}-\cite{Ranyuk}.


\begin{thebibliography}{9}
\bibitem{DVbook}
Dmitry Vasilievich Volkov. Papers, essays, reminiscences /  S.I.~Volkova, A.Yu.~Nurmagambetov. Kharkov, 2007. - 706p. ISBN 978-966-8661-25-9 (in Russian).
\bibitem{DVbrochure}
Dmitry Vasilievich Volkov / NSC KIPT, S.I.~Volkova. Kharkov, 2000. - 60p., photo. ISBN 966-7706-03-2 (in Russian).
\bibitem{sugra}
D.V.~Volkov, Supergravity before 1976. Proc. Intl. Conf. on the History of Original Ideas and Basic Discoveries in Particle Physics (Erice, Italy, 1994) published by Plenum Publishing Co. 1996, pp.663-674.
\bibitem{gap}
D.V.~Volkov, Generalized action principle for superstrings and supermembranes. Proc. Intl. Workshop on Supersymmetry and Unification of Fundamental Interactions SUSY'95 (Paris, France, 1995) published by Editions Frontieres 1996, pp.509-518.
\bibitem{SIVolkova}
S.I.~Volkova and A.A.~Zheltukhin, Glimpses of Dmitry Volkov's life and work, Nucl. Phys. B Proc. Suppl. \textbf{101} (2001) 20-25.
\bibitem{Zheltukhin:2009js}
 A.A.~Zheltukhin, Dmitrij Volkov, super-Poincare group and Grassmann variables, Annalen Phys.  \textbf{19} (2010) 177-185, \href{http://arxiv.org/abs/0911.0550}{arxiv:0911.0550 [hep-th]}. 
\bibitem{Duplij}
Memorial webpage of D.V.~Volkov is maintained by Steven Duplij at \href{http://homepages.spa.umn.edu/~duplij/volkov/}{http://homepages.spa.umn.edu/$\sim$duplij/volkov/}.
\bibitem{Ranyuk}
Oral interview with D.V.~Volkov by Yu.N.~Ranyuk is stored in the Niels Bohr Library of American Institute of Physics at  \href{http://www.aip.org/history-programs/niels-bohr-library/oral-histories/4392}{http://www.aip.org/history-programs/niels-bohr-library/oral-histories/4392}.
\end{thebibliography}
\end{document}